\documentclass[10pt, conference, letterpaper]{IEEEtran-v17}

\usepackage{xspace}
\usepackage{url}
\usepackage[cmex10]{amsmath}
\usepackage{bbm}
\usepackage{graphicx}
\usepackage{epstopdf}
\usepackage{paralist}
\usepackage[normalem]{ulem}
\usepackage{fancyref} 
\usepackage{amsmath}
\usepackage{amssymb}
\usepackage[stretch=16,shrink=16,step=4]{microtype}
\usepackage[table]{xcolor}
\usepackage{xcolor}
\definecolor{links}{rgb}{0.7,0,0}   
\definecolor{urls}{rgb}{0,0,0.8}    
\definecolor{cites}{rgb}{0,0,0.8}   
\usepackage{dsfont}
\usepackage{footnote}
\usepackage{balance}
\usepackage{float}
\usepackage{cite}
\usepackage{wasysym}
\usepackage{amssymb}

\usepackage{tabularx,booktabs}
\newcolumntype{L}[1]{>{\raggedright\let\newline\\\arraybackslash\hspace{0pt}}m{#1}}
\newcolumntype{C}[1]{>{\centering\let\newline\\\arraybackslash\hspace{0pt}}m{#1}}
\newcolumntype{R}[1]{>{\raggedleft\let\newline\\\arraybackslash\hspace{0pt}}m{#1}}
\usepackage{boldline}

\usepackage{vmr-symbols-vecbold}
\usepackage{standard-macros}
\usepackage{mathbbol}

\usepackage{tikz}
\usetikzlibrary{patterns}
\usetikzlibrary{shapes,arrows,positioning}
\usetikzlibrary{calc}
\usetikzlibrary{fit}
\usetikzlibrary{plotmarks}
\tikzset{every picture/.style={font issue=\scriptsize, >=stealth},font issue/.style={execute at begin picture={#1\selectfont}}}
\tikzset{three sided left/.style={
        draw=none,
        xshift=\pgflinewidth,
        append after command={
            [shorten <= -0.5\pgflinewidth]
            ([shift={(-1.5\pgflinewidth,-0.5\pgflinewidth)}]\tikzlastnode.north east) edge ([shift={( 0.5\pgflinewidth,-0.5\pgflinewidth)}]\tikzlastnode.north west) 
            ([shift={( 0.5\pgflinewidth,-0.5\pgflinewidth)}]\tikzlastnode.north west) edge ([shift={( 0.5\pgflinewidth,+0.5\pgflinewidth)}]\tikzlastnode.south west)            
            ([shift={( 0.5\pgflinewidth,+0.5\pgflinewidth)}]\tikzlastnode.south west) edge ([shift={(-1.0\pgflinewidth,+0.5\pgflinewidth)}]\tikzlastnode.south east)
        }}}
        
\tikzset{three sided right/.style={
        draw=none,
        xshift=-\pgflinewidth,
        append after command={
            [shorten <= -0.5\pgflinewidth]
            ([shift={( 1.5\pgflinewidth,-0.5\pgflinewidth)}]\tikzlastnode.north west) edge ([shift={(-0.5\pgflinewidth,-0.5\pgflinewidth)}]\tikzlastnode.north east) 
            ([shift={(-0.5\pgflinewidth,-0.5\pgflinewidth)}]\tikzlastnode.north east) edge ([shift={(-0.5\pgflinewidth,+0.5\pgflinewidth)}]\tikzlastnode.south east)            
            ([shift={(-0.5\pgflinewidth,+0.5\pgflinewidth)}]\tikzlastnode.south east) edge ([shift={( 1.0\pgflinewidth,+0.5\pgflinewidth)}]\tikzlastnode.south west)
        }}}

\usepackage{pgfplots}
\usepgfplotslibrary{fillbetween}
\pgfplotsset{
  compat=newest, 
  width=\columnwidth,    
  height=0.8\columnwidth,   
  plot coordinates/math parser=false,
  standard/.style={
    axis equal,
    axis line style=help lines,
    axis x line=center,
    axis y line=center,
    axis z line=center},
    grid style={dashed,gray},
    minor grid style={dotted,gray},
    major grid style={dotted,gray},
    ylabel absolute, ylabel style={yshift=-0.4cm},
    xlabel absolute, xlabel style={yshift=0.25cm}
}

\makeatletter
 \pgfdeclarepatternformonly[\tikz@pattern@color,\LineSpace,\LineWidth]{my horizontal lines}%
    {\pgfpointorigin}{\pgfqpoint{100pt}{1pt}}{\pgfqpoint{100pt}{\LineSpace}}%
    {
        \pgfsetcolor{\tikz@pattern@color}
        \pgfsetlinewidth{\LineWidth}
        \pgfpathmoveto{\pgfqpoint{0pt}{0.5pt}}
        \pgfpathlineto{\pgfqpoint{100pt}{0.5pt}}
        \pgfusepath{stroke}
    }
 
 \pgfdeclarepatternformonly[\tikz@pattern@color,\LineSpace,\LineWidth]{my vertical lines}%
    {\pgfpointorigin}{\pgfqpoint{1pt}{100pt}}{\pgfqpoint{\LineSpace}{100pt}}%
    {
        \pgfsetcolor{\tikz@pattern@color}
        \pgfsetlinewidth{\LineWidth}
        \pgfpathmoveto{\pgfqpoint{0.5pt}{0pt}}
        \pgfpathlineto{\pgfqpoint{0.5pt}{100pt}}
        \pgfusepath{stroke}
    } 
 
 \pgfdeclarepatternformonly[\tikz@pattern@color,\LineSpace,\LineWidth]{my grid}%
    {\pgfqpoint{-1pt}{-1pt}}{\pgfqpoint{\LineSpace}{\LineSpace}}
    {\pgfqpoint{\LineSpace}{\LineSpace}}%
    {
        \pgfsetcolor{\tikz@pattern@color}
        \pgfsetlinewidth{\LineWidth}
        \pgfpathmoveto{\pgfqpoint{0pt}{0pt}}
        \pgfpathlineto{\pgfqpoint{0pt}{\LineSpace + 0.1pt}}
        \pgfpathmoveto{\pgfqpoint{0pt}{0pt}}
        \pgfpathlineto{\pgfqpoint{\LineSpace + 0.1pt}{0pt}}
        \pgfusepath{stroke}
    }
 
 \pgfdeclarepatternformonly[\tikz@pattern@color,\LineSpace,\LineWidth]{my north east lines}
    {\pgfqpoint{-\LineWidth}{-\LineWidth}}{\pgfqpoint{\LineSpace}{\LineSpace}}
    {\pgfqpoint{\LineSpace}{\LineSpace}}%
    {
        \pgfsetcolor{\tikz@pattern@color}
        \pgfsetlinewidth{\LineWidth}
        \pgfpathmoveto{\pgfqpoint{-\LineWidth}{-\LineWidth}}
        \pgfpathlineto{\pgfqpoint{\LineSpace + 0.1pt}{\LineSpace + 0.1pt}}
        \pgfusepath{stroke}
    }
 
 \pgfdeclarepatternformonly[\tikz@pattern@color,\LineSpace,\LineWidth]{my north west lines}
    {\pgfqpoint{-\LineWidth}{-\LineWidth}}{\pgfqpoint{\LineSpace}{\LineSpace}}
    {\pgfqpoint{\LineSpace}{\LineSpace}}%
    {
        \pgfsetcolor{\tikz@pattern@color}
        \pgfsetlinewidth{\LineWidth}
        \pgfpathmoveto{\pgfqpoint{-\LineWidth}{\LineSpace}}
        \pgfpathlineto{\pgfqpoint{\LineSpace + 0.1pt}{-\LineWidth}}
        \pgfusepath{stroke}
    }
 
 \pgfdeclarepatternformonly[\tikz@pattern@color,\LineSpace,\LineWidth]{my crosshatch}%
    {\pgfqpoint{-1pt}{-1pt}}{\pgfqpoint{\LineSpace}{\LineSpace}}
    {\pgfqpoint{\LineSpace}{\LineSpace}}%
    {
        \pgfsetcolor{\tikz@pattern@color}
        \pgfsetlinewidth{\LineWidth}
        \pgfpathmoveto{\pgfqpoint{\LineSpace + 0.1pt}{0pt}}
        \pgfpathlineto{\pgfqpoint{0pt}{\LineSpace + 0.1pt}}
        \pgfpathmoveto{\pgfqpoint{0pt}{0pt}}
        \pgfpathlineto{\pgfqpoint{\LineSpace + 0.1pt}{\LineSpace + 0.1pt}}
        \pgfusepath{stroke}
    }
 
 \pgfdeclarepatternformonly[\tikz@pattern@color,\LineSpace,\PointSize]{my dots}%
    {\pgfqpoint{-\LineSpace*0.25}{-\LineSpace*0.25}}
    {\pgfqpoint{\LineSpace*0.25}{\LineSpace*0.25}}
    {\pgfqpoint{\LineSpace*0.75}{\LineSpace*0.75}}%
    {
        \pgfsetcolor{\tikz@pattern@color}
        \pgfpathcircle{\pgfqpoint{0pt}{0pt}}{\PointSize}
        \pgfusepath{fill}
    }
\makeatother
 
\newdimen\LineSpace
\newdimen\PointSize
\newdimen\LineWidth
\tikzset{
    line space/.code={\LineSpace=#1},
    line space=3pt
}
\tikzset{
    point size/.code={\PointSize=#1},
    point size=.5pt
}
\tikzset{
    pattern line width/.code={\LineWidth=#1},
    pattern line width=.4pt
}

\DeclareSymbolFontAlphabet{\amsmathbb}{AMSb}%

\newcommand{\lro}[1]{\lefto({#1}\right)}																
\newcommand{\lrbo}[1]{\lefto \lbrace {#1} \right \rbrace}															
\newcommand{\lrho}[1]{\lefto [ {#1} \right ]}																				

\newcommand{\lr}[1]{\left({#1}\right)}																
\newcommand{\lrb}[1]{\left \lbrace {#1} \right \rbrace}															

\safemath{\dopplerspread}{B_D}																								
\safemath{\delayspread}{T_D}																									
\safemath{\nc}{n\sub{c}}																										
\safemath{\nf}{n\sub{f}}																										
\safemath{\efa}{\epsilon\sub{a}}
\safemath{\efb}{\epsilon\sub{b}}
\safemath{\ef}{\epsilon\sub{f}	}
\safemath{\nd}{n\sub{d}}																										
\safemath{\ntx}{n\sub{t}} 																											
\safemath{\nrx}{n\sub{r}}																											
\safemath{\ntxt}{\tilde{n\sub{t}}}																											
\safemath{\cb}{\ensuremath{L}} 																								
\safemath{\cl}{\ensuremath{n}} 																								
\safemath{\txanto}{{\ensuremath{\tilde{m}_t}}} 																		
\safemath{\cs}{M} 																														
\safemath{\idPustm}{\ensuremath{S_{k}}}
\safemath{\error}{\ensuremath{\epsilon}} 																				
\safemath{\eexp}{\ensuremath{\mathcal{E}}} 																			
\safemath{\nsubc}{n\sub{s}}			 																						
\safemath{\nofdm}{n\sub{o}} 																									
\safemath{\bc}{\ensuremath{B_c}} 																							
\safemath{\ts}{\ensuremath{T_s}} 																							
\safemath{\nrb}{\ensuremath{n_{rb}}} 																						
\safemath{\nres}{\ell}
\safemath{\maxk}{M^*\lr{\nres, \nsubc, \nofdm, \epsilon, \rho}}
\safemath{\Rmax}{R^*}
\safemath{\Emin}{E\sub{b}^*/N_0}
\safemath{\Eminf}{\frac{E\sub{b}^*}{N_0}}
\safemath{\np}{\ensuremath{n\sub{p}}}
\safemath{\code}{\ensuremath{\mathcal{C}}}
\safemath{\err}{\ensuremath{\epsilon}}
\safemath{\rp}{\ensuremath{\rho\sub{p}}}
\safemath{\rd}{\ensuremath{\rho\sub{d}}}
\safemath{\cohtime}{\ensuremath{T\sub{c}}}
\safemath{\cohbw}{\ensuremath{B\sub{c}}}
\safemath{\nmax}{\ensuremath{\ell\sub{max}}}

\safemath{\yp}{\ensuremath{\randvecy^{(\text{p})}}}
\safemath{\yd}{\ensuremath{\randvecy^{(\text{d})}}}

\safemath{\xp}{\ensuremath{\vecx^{(\text{p})}}}
\safemath{\xd}{\ensuremath{\randvecx^{(\text{d})}}}
\safemath{\xpbar}{\ensuremath{\overline{\vecx}^{(\text{p})}}}
\safemath{\xdbar}{\ensuremath{\overline{\randvecx}^{(\text{d})}}}

\safemath{\xdv}{\ensuremath{\randvecx^{(\text{d})}}}
\safemath{\xdbarv}{\ensuremath{\overline{\randvecx}^{(\text{d})}}}
\safemath{\ydv}{\ensuremath{\randvecy^{(\text{d})}}}

\safemath{\xdr}{\ensuremath{\matX^{(\text{d})}}}

\safemath{\Pv}{\ensuremath{P\sub{av}\lro{a}}}

\newcommand{\prob}[1]{\ensuremath{\mathbb{P}\lrho{#1}}}

\safemath{\mI}{\ensuremath{i\lro{\randvecy ; \randvecx}}} 				

\safemath{\randveca}{\bm{A}}
\safemath{\randvecb}{\bm{B}}
\safemath{\randvecc}{\bm{C}}
\safemath{\randvecd}{\bm{D}}
\safemath{\randvece}{\bm{E}}
\safemath{\randvecf}{\bm{F}}
\safemath{\randvecg}{\bm{G}}
\safemath{\randvech}{\bm{H}}
\safemath{\randveci}{\bm{I}}
\safemath{\randvecj}{\bm{J}}
\safemath{\randveck}{\bm{K}}
\safemath{\randvecl}{\bm{L}}
\safemath{\randvecm}{\bm{M}}
\safemath{\randvecn}{\bm{N}}
\safemath{\randveco}{\bm{O}}
\safemath{\randvecp}{\bm{P}}
\safemath{\randvecq}{\bm{Q}}
\safemath{\randvecr}{\bm{R}}
\safemath{\randvecs}{\bm{S}}
\safemath{\randvect}{\bm{T}}
\safemath{\randvecu}{\bm{U}}
\safemath{\randvecv}{\bm{V}}
\safemath{\randvecw}{\bm{W}}
\safemath{\randvecx}{\bm{X}}
\safemath{\randvecy}{\bm{Y}}
\safemath{\randvecz}{\bm{Z}}
\safemath{\randvecphi}{\bm{\Phi}}

\safemath{\randmatA}{\amsmathbb{A}}
\safemath{\randmatB}{\amsmathbb{B}}
\safemath{\randmatC}{\amsmathbb{C}}
\safemath{\randmatD}{\amsmathbb{D}}
\safemath{\randmatE}{\amsmathbb{E}}
\safemath{\randmatF}{\amsmathbb{F}}
\safemath{\randmatG}{\amsmathbb{G}}
\safemath{\randmatH}{\amsmathbb{H}}
\safemath{\randmatI}{\amsmathbb{I}}
\safemath{\randmatJ}{\amsmathbb{J}}
\safemath{\randmatK}{\amsmathbb{K}}
\safemath{\randmatL}{\amsmathbb{L}}
\safemath{\randmatM}{\amsmathbb{M}}
\safemath{\randmatN}{\amsmathbb{N}}
\safemath{\randmatO}{\amsmathbb{O}}
\safemath{\randmatP}{\amsmathbb{P}}
\safemath{\randmatQ}{\amsmathbb{Q}}
\safemath{\randmatR}{\amsmathbb{R}}
\safemath{\randmatS}{\amsmathbb{S}}
\safemath{\randmatT}{\amsmathbb{T}}
\safemath{\randmatU}{\amsmathbb{U}}
\safemath{\randmatV}{\amsmathbb{V}}
\safemath{\randmatW}{\amsmathbb{W}}
\safemath{\randmatX}{\amsmathbb{X}}
\safemath{\randmatY}{\amsmathbb{Y}}
\safemath{\randmatZ}{\amsmathbb{Z}}
\safemath{\randmatSigma}{\mathbb{\Sigma}}
\safemath{\randmatPhi}{\mathbb{\Phi}}
\safemath{\randmatLambda}{\mathbb{\Lambda}}

\safemath{\matSigma}{\bm{\Sigma}}
\safemath{\matPhi}{\bm{\Phi}}
\safemath{\matLambda}{\bm{\Lambda}}

\usepackage{glossaries}
\makeglossaries
\newglossaryentry{paoi}
{
  name={PAoI},
  description={peak age of information},
  first={\glsentrydesc{paoi} (\glsentrytext{paoi})}
}

\newglossaryentry{pgf}
{
  name={PGF},
  description={probability-generating function},
  first={\glsentrydesc{pgf} (\glsentrytext{pgf})}
}

\newglossaryentry{biawgn}
{
  name={BI-AWGN},
  description={binary-input additive white Gaussian noise},
  first={\glsentrydesc{biawgn} (\glsentrytext{biawgn})}
}
\newglossaryentry{bsc}
{
  name={BSC},
  description={binary-symmetric channel},
  first={\glsentrydesc{bsc} (\glsentrytext{bsc})}
}

\newglossaryentry{bec}
{
  name={BEC},
  description={binary-erasure channel},
  first={\glsentrydesc{bec} (\glsentrytext{bec})}
}
\newglossaryentry{awgn}
{
  name={AWGN},
  description={additive white Gaussian noise},
  first={\glsentrydesc{awgn} (\glsentrytext{awgn})}
}
\newglossaryentry{3gpp}
{
  name={3GPP},
  description={3rd generation partnership project},
  first={\glsentrydesc{3gpp} (\glsentrytext{3gpp})}
}

\newglossaryentry{dl}
{
  name={DL},
  description={downlink},
  first={\glsentrydesc{dl} (\glsentrytext{dl})}
}

\newglossaryentry{LED}
{
  name={LED},
  description={light emitting diode},
  first={\glsentrydesc{LED} (\glsentrytext{LED})},
  plural={LEDs},
  descriptionplural={light emitting diodes},
  firstplural={\glsentrydescplural{LED} (\glsentryplural{LED})}
}

\newglossaryentry{lte}
{
  name={LTE},
  description={long term evolution},
  first={\glsentrydesc{lte} (\glsentrytext{lte})}
}
\newglossaryentry{ofdm}
{
  name={OFDM},
  description={orthogonal frequency-division multiplexing},
  first={\glsentrydesc{ofdm} (\glsentrytext{ofdm})}
}
\newglossaryentry{ue}
{
  name={UE},
  description={user equipment},
  first={\glsentrydesc{ue} (\glsentrytext{ue})},
  plural={UE's},
  descriptionplural={user equipments},
  firstplural={\glsentrydescplural{ue} (\glsentryplural{ue})}
}
\newglossaryentry{ul}
{
  name={UL},
  description={uplink},
  first={\glsentrydesc{ul} (\glsentrytext{ul})}
}
\newglossaryentry{rb}
{
  name={RB},
  description={resource block},
  first={\glsentrydesc{rb} (\glsentrytext{rb})},
  plural={RBs},
  descriptionplural={resource blocks},
  firstplural={\glsentrydescplural{rb} (\glsentryplural{rb})}
}
\newglossaryentry{cu}
{
  name={CU},
  description={channel use},
  first={\glsentrydesc{cu} (\glsentrytext{cu})},
  plural={CUs},
  descriptionplural={channel uses},
  firstplural={\glsentrydescplural{cu} (\glsentryplural{cu})}
}
\newglossaryentry{tti}
{
  name={TTI},
  description={transmission time interval},
  first={\glsentrydesc{tti} (\glsentrytext{tti})},
  plural={TTI's},
  descriptionplural={transmission time intervals},
  firstplural={\glsentrydescplural{tti} (\glsentryplural{tti})}
}
\newglossaryentry{iid}
{
  name={i.i.d.},
  description={identical and independently distributed},
  first={\glsentrydesc{iid} (\glsentrytext{iid})}
}
\newglossaryentry{psd}
{
  name={PSD},
  description={power spectral density},
  first={\glsentrydesc{psd} (\glsentrytext{psd})}
}

\newglossaryentry{mimo}
{
  name={MIMO},
  description={multiple-input multiple-output},
  first={\glsentrydesc{mimo} (\glsentrytext{mimo})}
}

\newglossaryentry{bler}
{
  name={BLER},
  description={block error probability},
  first={\glsentrydesc{bler} (\glsentrytext{bler})}
}
\newglossaryentry{mtc}
{
  name={MTC},
  description={machine-type communication},
  first={\glsentrydesc{mtc} (\glsentrytext{mtc})}
}
\newglossaryentry{csi}
{
  name={CSI},
  description={channel state information},
  first={\glsentrydesc{csi} (\glsentrytext{csi})}
}
\newglossaryentry{dvbt}
{
  name={DVB-T},
  description={terrestrial digital video broadcasting},
  first={\glsentrydesc{dvbt} (\glsentrytext{dvbt})}
}
\newglossaryentry{pat}
{
  name={PAT},
  description={pilot-assisted transmission},
  first={\glsentrydesc{pat} (\glsentrytext{pat})}
}
\newglossaryentry{ml}
{
  name={ML},
  description={maximum likelihood},
  first={\glsentrydesc{ml} (\glsentrytext{ml})}
}
\newglossaryentry{dt}
{
  name={DT},
  description={dependence-testing},
  first={\glsentrydesc{dt} (\glsentrytext{dt})}
}
\newglossaryentry{ustm}
{
  name={USTM},
  description={unitary space-time modulation},
  first={\glsentrydesc{ustm} (\glsentrytext{ustm})}
}
\newglossaryentry{siso}
{
  name={SISO},
  description={single-input single-output},
  first={\glsentrydesc{siso} (\glsentrytext{siso})}
}
\newglossaryentry{snn}
{
  name={SNN},
  description={scaled nearest-neighbor},
  first={\glsentrydesc{snn} (\glsentrytext{snn})}
}
\newglossaryentry{rcus}
{
  name={RCUs},
  description={random-coding union bound with parameter $s$},
  first={\glsentrydesc{rcus} (\glsentrytext{rcus})}
}
\newglossaryentry{osd}
{
  name={OSD},
  description={ordered statistics decoding},
  first={\glsentrydesc{osd} (\glsentrytext{osd})}
}
\newglossaryentry{llr}
{
  name={LLR},
  description={log-likelihood ratio},
  first={\glsentrydesc{llr} (\glsentrytext{llr})}
   plural={LLR's},
  descriptionplural={log-likelihood ratios},
  firstplural={\glsentrydescplural{llr} (\glsentryplural{llr})}
}
\newglossaryentry{qpsk}
{
  name={QPSK},
  description={quaternary phase shift keying},
  first={\glsentrydesc{qpsk} (\glsentrytext{qpsk})}
}
\newglossaryentry{nlos}
{
  name={NLOS},
  description={non-line-of-sight},
  first={\glsentrydesc{nlos} (\glsentrytext{nlos})}
}
\newglossaryentry{los}
{
  name={LOS},
  description={line-of-sight},
  first={\glsentrydesc{los} (\glsentrytext{los})}
}
\newglossaryentry{cgf}
{
  name={CGF},
  description={cumulant-generating function},
  first={\glsentrydesc{cgf} (\glsentrytext{cgf})}
   plural={CGF's},
  descriptionplural={cumulant-generating functions},
  firstplural={\glsentrydescplural{cgf} (\glsentryplural{cgf})}
}
\newglossaryentry{pdf}
{
  name={PDF},
  description={probability density function},
  first={\glsentrydesc{pdf} (\glsentrytext{pdf})}
   plural={PDF's},
  descriptionplural={probability density functions},
  firstplural={\glsentrydescplural{pdf} (\glsentryplural{pdf})}
}
\newglossaryentry{gmi}
{
  name={GMI},
  description={generalized mutual information},
  first={\glsentrydesc{gmi} (\glsentrytext{gmi})}
}

\newglossaryentry{arq}
{
  name={ARQ},
  description={automatic repeat request},
  first={\glsentrydesc{arq} (\glsentrytext{arq})}
}

\newglossaryentry{harq}
{
  name={HARQ},
  description={hybrid automatic repeat request},
  first={\glsentrydesc{harq} (\glsentrytext{harq})}
}

\newglossaryentry{fbl}
{
  name={FBL-NF},
  description={fixed blocklength with no feedback},
  first={\glsentrydesc{fbl} (\glsentrytext{fbl})}
}
\newglossaryentry{ssnf}
{
  name={SS-NF},
  description={single-shot no-feedback},
  first={\glsentrydesc{ssnf} (\glsentrytext{ssnf})}
}

\newglossaryentry{urllc}
{
  name={URLLC},
  description={ultra-reliable low-latency communications},
  first={\glsentrydesc{urllc} (\glsentrytext{urllc})}
}

\newglossaryentry{vlsf}
{
  name={VLSF},
  description={variable-length stop-feedback},
  first={\glsentrydesc{vlsf} (\glsentrytext{vlsf})}
}

\newglossaryentry{dmt}
{
  name={DMT},
  description={diversity-multiplexing tradeoff},
  first={\glsentrydesc{dmt} (\glsentrytext{dmt})}
}

\newglossaryentry{dmdt}
{
  name={DMDT},
  description={diversity-multiplexing-delay tradeoff},
  first={\glsentrydesc{dmdt} (\glsentrytext{dmdt})}
}

\newglossaryentry{tddt}
{
  name={TDDT},
  description={throughput-diversity-delay tradeoff},
  first={\glsentrydesc{tddt} (\glsentrytext{tddt})}
}

\newglossaryentry{stbc}
{
  name={STBC},
  description={space-time block-codes},
  first={\glsentrydesc{stbc} (\glsentrytext{stbc})},
  plural={STBC's},
  descriptionplural={space-time block-codes},
  firstplural={\glsentrydescplural{stbc} (\glsentryplural{stbc})}
}

\newglossaryentry{harqir}
{
  name={HARQ-IR},
  description={hybrid automatic repeat request with incremental redundancy},
  first={\glsentrydesc{harqir} (\glsentrytext{harqir})}
}

\newglossaryentry{harqcc}
{
  name={HARQ-CC},
  description={hybrid automatic repeat request with chase combining},
  first={\glsentrydesc{harqcc} (\glsentrytext{harqcc})}
}

\newglossaryentry{wp1}
{
  name={w.p.1.},
  description={with probability one},
  first={\glsentrydesc{wp1} (\glsentrytext{wp1})}
}

\newglossaryentry{vlf}
{
  name={VLF},
  description={variable-length feedback},
  first={\glsentrydesc{vlf} (\glsentrytext{vlf})}
}

\newglossaryentry{dmc}
{
  name={DMC},
  description={discrete memoryless channel},
  first={\glsentrydesc{dmc} (\glsentrytext{dmc})}
}

\usepackage{pgfplots}
\usepgfplotslibrary{groupplots}
\usetikzlibrary{pgfplots.groupplots}

\usepgfplotslibrary{external}

\newcommand\linew{1pt} 
\newcommand{\fwidth}{\textwidth}

\makeatletter
\def\@IEEEinterspaceratioM{0.265}
\def\@IEEEinterspaceMINratioM{0.1651}
\def\@IEEEinterspaceMAXratioM{0.38}

\def\@IEEEinterspaceratioB{0.31}
\def\@IEEEinterspaceMINratioB{0.19}
\def\@IEEEinterspaceMAXratioB{0.38}
\@IEEEtunefonts
\makeatother
\definecolor{red}{RGB}{220, 10, 10}
\definecolor{blue}{RGB}{10, 40, 200}
\definecolor{green}{RGB}{10, 200, 10}
\definecolor{orange}{RGB}{255, 165, 0}
\definecolor{pink}{RGB}{255, 51, 204}

\pgfplotscreateplotcyclelist{colorbrewer}{
{red},
{blue},
{green},
{orange},
{pink}
}
\pgfplotscreateplotcyclelist{colorfour}{
{red},
{blue},
{green},
{orange}
}

\usepackage[font=scriptsize]{subcaption}
\let\marginnote\undefined

\newcommand{\marginnote}[1]{\marginpar{\framebox{\framebox{#1}}}}

\allowdisplaybreaks
\begin{document}
\IEEEoverridecommandlockouts
\title{Peak-Age Violation Guarantees for the Transmission of Short Packets over Fading Channels}
\author{\IEEEauthorblockN{Johan \"Ostman$^1$, Rahul Devassy$^1$, Giuseppe Durisi$^1$, and Elif Uysal$^2$\\
$^1$Chalmers University of Technology,  Gothenburg, Sweden \\
$^2$Middle East Technical University, Ankara, Turkey}
\thanks{This work was partly supported by the Swedish Research Council under grants 2014-6066 and 2016-03293, and by TUBITAK grant 117E215.}
}
\maketitle

\begin{abstract}
  We investigate the probability that the peak age of information in a point-to-point communication system operating over a  multiantenna wireless fading channel exceeds a predetermined value.
The packets are scheduled according to a last-come first-serve policy with preemption in service, and are transmitted over the channel using a simple automatic repetition request protocol.
We consider quadrature phase shift keying modulation, pilot-assisted transmission, maximum-likelihood channel estimation, and mismatched scaled nearest-neighbor decoding.
Our analysis, which exploits nonasymptotic tools in information theory, allows one to determine, for a given information packet size, the physical layer parameters such as the SNR, the number of transmit and receive antennas, the amount of frequency diversity to exploit, and the number of pilot symbols, to ensure that the system operates below a target peak-age violation probability.
\end{abstract}

\section{Introduction}
Emerging machine-type applications such as industrial automation and control, traffic safety through automated transportation, and tactile internet, require the availability of wireless communication systems that can exchange short information packets under stringent latency and reliability constraints~\cite{durisi16-09a}.
Supporting such a use case, commonly referred to as \gls{urllc}, is a key objective of the upcoming next-generation wireless cellular system (5G).

The design of \gls{urllc} systems operating over wireless fading channels presents many challenges at the physical layer:
\begin{inparaenum}[i)]
  \item information packets may be generated sporadically, which implies that \gls{csi} is typically not available at transmitter and receiver;
  \item frequency diversity and spatial diversity through the use of multiple antennas need to be exploited simultaneously to achieve the desired reliability levels~\cite{durisi16-02a,ferrante18,johansson15-06a};
  \item since latency constraints impose the transmission of short coded packets, one needs to account for finite blocklength effects in performance analyses~\cite{polyanskiy10-05a}, by using more sophisticated tools than the classical outage and ergodic capacity formulas~\cite{durisi16-09a}.
\end{inparaenum}

The packet size is, however, not the only factor that determines the communication latency.
Another crucial factor is the queuing delay resulting from the presence of a data stream.
An important performance metric in \gls{urllc} is then the probability that the latency of a packet, including both coding and queuing delay exceeds a given value.
Such metric, which we shall refer to as delay violation probability, has been recently characterized in~\cite{schiessl18-12a} using a network-calculus upper bound~\cite{al-zubaidy16-02a} in the context of short-packet transmission over fading channels in the presence of imperfect \gls{csi} at the transmitter.
Furthermore, an exact characterization of this metric was recently provided in~\cite{devassy18-06a} in the context of transmission over a nonfading AWGN channel.
The queuing-delay violation probability has been upper-bounded using the effective bandwidth framework~\cite{chang95-08a} in the context of downlink multiuser transmission~\cite{she16-12b} and the effective-capacity framework~\cite{wu03-02a} in the context of transmission over point-to-point fading channels~\cite{gursoy13-12a}.

However, in some \gls{urllc} applications, minimizing the packet delay violation probability, although relevant from a link-layer viewpoint, may be misaligned with the actual requirements at the application layer.
Consider for example the case of factory automation.
There, the information packets exchanged over the wireless medium may carry sensor data used to track a remote process at a given destination.
The relevant performance metric in such a scenario is the freshness of the sensor data available at the destination, rather than the delay of each packet~\cite{devassy18-12a}.
Indeed, packets that contain outdated sensor data are not of interest to the destination and should simply be dropped rather than delivered with low latency and high reliability.

In such a context, a more relevant performance metric is the probability that the \emph{peak age of information} (PAoI), which describes the maximum time that is elapsed since the last update was received at the destination (see, e.g,~\cite{Costa16-04}), exceeds a predetermined value.
We shall refer to this quantity as PAoI violation probability.

While the delay violation probability may be equivalent to the PAoI violation probability under the assumption of a constant flow of packets, these two metrics can differ drastically in other cases. For example, when packet transmissions are too infrequent, there will be aging even when  queuing and coding delays are negligible. On the contrary, when packet transmissions are subject to significant random delays, the throughput needs to be limited in order to achieve a low age violation probability~\cite{Sun17-11}.

The steady-state distribution of the peak age of information (from which the PAoI violation probability can be obtained) was characterized in~\cite{Costa16-04} for M/M/1 queues with system capacity $1$ and $2$ with and without preemption in the queue.
More general queuing models, under the simplifying assumption that all packets are informative and need to be delivered, were considered in~\cite{inoue17-06a}.
These two studies, however, used an abstract model for the service process, which is too crude to capture the coding aspects for packet transmission over wireless channels.
Recently,~\cite{devassy18-12a} analyzed PAoI and delay violation probability under a more accurate model of channel coding.
This model takes advantage of recent results in finite-blocklength information theory~\cite{polyanskiy10-05a, martinez11-02a} for general \emph{memoryless} channels.
Furthermore, a new packet management scheme, which is based on a last-come first-serve discipline with preemption in service (LCFS-S), is analyzed.
For the case in which packets are transmitted using a simple \gls{arq} protocol, this policy is shown to outperform the three policies studied in~\cite{Costa16-04} in terms of PAoI violation probability.

\paragraph*{Contributions}
{By combining the results in \cite{ferrante18} and \cite{devassy18-12a}, we perform an analysis of the AoI violation probability in a practical wireless communication system.}
In particular, we extend the analysis of the PAoI violation probability under LCFS-S and simple \gls{arq} transmission in~\cite{devassy18-12a} by dropping the memoryless channel assumption.
Specifically, we model the propagation channel as a \gls{mimo}, spatially white, Rayleigh block-fading channel.
As discussed in~\cite{ferrante18}, this model is relevant for 5G because it allows one to capture both frequency, time, and spatial diversity.
Indeed, in 5G, a codeword may be divided into resource blocks that are transmitted on different diversity branches, both in time, and frequency.
Exploiting frequency diversity instead of time diversity is typically preferable under latency constraints.

We consider the practically relevant case of quadrature phase shift keying (QPSK) transmission, which is suitable for URLLC because of the low data rate.
Furthermore, we assume pilot-assisted transmission, maximum-likelihood channel estimation at the receiver, and a simple mismatched scaled nearest-neighbor decoder that treats the estimated \gls{csi} as if it was perfect.

For a given number of information bits and a given coding rate, our analysis allows one to determine the physical layer parameters, i.e., number of antennas at the transmitter and the receiver, the number of frequency diversity branches to code over, the number of pilot symbols, and the SNR, that are needed to not exceed a target PAoI violation probability.

\paragraph*{Notation}
Scalar quantities are denoted by normal font letters whereas vectors and matrices are denoted by lowercase and uppercase boldface letters, respectively.
We denote the distribution of a circularly-symmetric complex Gaussian random variable with zero mean and variance $\sigma^2$ by $\jpg(0,\sigma^2)$ and by $\mathrm{Geom}(p)$ the distribution of a geometrically distributed random variable with parameter $p$.
The superscript $\herm{\lro{\cdot}}$ denotes Hermitian transposition and we write $\log\lro{\cdot}$ to denote the natural logarithm.
Finally, $\lrho{a}^+$ stands for $\max\lrbo{0,a}$, $\vecnorm{\cdot}$ for the $\ell_2$ norm, $\Ex{}{\cdot}$ for the expectation operator, and $G_X\lro{s} = \Ex{}{s^X}$ for the \gls{pgf} of a nonnegative integer random variable $X$.

\section{System Model}
\label{sec:system_model}
We consider a discrete-time setup in which time is organized in channel uses over which a single complex-valued coded symbol may be transmitted from each transmit antenna over the wireless channel.
We assume that, in each  channel use, an information packet consisting of $k$ information bits reaches the transmitter queue with probability $\lambda$.
Hence, $1/\lambda$ is the average interarrival time between successive information packets.

\begin{figure}[t]
        \centering
          \def \bufferlength{0.3}
\def \bufferwidth{0.3}
\def \buffery{-0.3}
\def \preempy{-2.5}

\begin{tikzpicture}[thick,scale=1, every node/.style={scale=1}]


\tikzset{
block/.style = {draw, fill=white, rectangle, minimum height=2em, minimum width=3em},
bufferblock/.style = {draw, fill=white, rectangle, minimum height=2cm, minimum width=1cm},
tmp/.style  = {coordinate},
sum/.style= {draw, fill=white, circle, node distance=1cm},
input/.style = {coordinate},
output/.style= {coordinate},
pinstyle/.style = {pin edge={to-,thin,black}},
queuei/.pic={
  \draw (0,0) -- ++(2cm,0) -- ++(0,-0.5cm) -- ++(-2cm,0);
   \foreach \c [count =\x from 1] in {$\randvecy_1$,$\cdots$,$\randvecy_{v}$}
   {
    \draw ([xshift=-\x*15pt]2cm,0) -- ++(0,-0.5cm) {};
    \node[text centered] at ([xshift=-\x*15pt]2.3cm,-0.25cm) {\c};
   }
   \node[above] at (1cm,0) {Buffer};
  },
  queuel/.pic={
  \draw (2cm,0) -- ++(-2cm,0) -- ++(0,-0.5cm) -- ++(2cm,0);
   \foreach \c [count =\x from 1] in {,,}
   {
    \draw ([xshift=-\x*15pt]2.1cm,0) -- ++(0,-0.5cm) {};
   }
  },
  bsc/.pic={
     \node[circle,fill=blue,inner sep=1pt,minimum size=1pt] (A1) at (0,0.5) {} ;
      \node[circle,fill=blue,inner sep=1pt,minimum size=1pt] (A2)  at (0,0) {} ;
     \node[circle,fill=blue,inner sep=1pt,minimum size=1pt] (B1) at (0.5,0.5) {} ;
      \node[circle,fill=blue,inner sep=1pt,minimum size=1pt] (B2)  at (0.5,0) {};
      \draw[-]  (A1)--(B1) node [midway, above, sloped]  {};
      \draw[-]  (A1)--(B2) node at (1,2.4) {};
      \draw[-]  (A2)--(B1) node  at (1,0.6) {};
      \draw[-]  (A2)--(B2) node [midway, below, sloped]  {};
  }
}

\draw[->] (0,0) -- +(5.5,0) node[anchor=west] (timea) {\scriptsize Channel};
\node[below = 0.25cm of timea.west, anchor = west]  {\scriptsize uses};

\foreach \x in {0,...,25} {
  \draw[-,>=stealth] (\x*0.2,-0.05) -- +(0,0.1);
}

\foreach \x in {0,5,10,15,20, 25} {
  \draw[-, thick] (\x*0.2,-0.1) -- +(0,0.2);
}

\foreach \x in {5,10,15} {
  \draw[draw=black, fill=orange!80!black] (\x*0.2,0.3) rectangle ++(1,0.3);
}

\draw (-0.3, \buffery) -- ++(\bufferlength,0) -- ++(0,-\bufferwidth) -- ++(-\bufferlength,0);
\draw (-0.15, \buffery) -- +(0,-\bufferwidth);
\draw (0.7, \buffery) -- ++(\bufferlength,0) -- ++(0,-\bufferwidth) -- ++(-\bufferlength,0);
\draw[draw=black, fill=orange!80!black] (0.85, \buffery) rectangle ++(\bufferlength/2,-\bufferwidth);
\draw (1.7, \buffery) -- ++(\bufferlength,0) -- ++(0,-\bufferwidth) -- ++(-\bufferlength,0);
\draw[draw=black, fill=orange!80!black] (1.85, \buffery) rectangle ++(\bufferlength/2,-\bufferwidth);
\draw (2.7, \buffery) -- ++(\bufferlength,0) -- ++(0,-\bufferwidth) -- ++(-\bufferlength,0);
\draw[draw=black, fill=orange!80!black] (2.85, \buffery) rectangle ++(\bufferlength/2,-\bufferwidth);
\draw (3.7, \buffery) -- ++(\bufferlength,0) -- ++(0,-\bufferwidth) -- ++(-\bufferlength,0);
\draw (3.85, \buffery) -- +(0,-\bufferwidth);
\draw (4.7, \buffery) -- ++(\bufferlength,0) -- ++(0,-\bufferwidth) -- ++(-\bufferlength,0);
\draw (4.85, \buffery) -- +(0,-\bufferwidth);

\draw[draw=black, fill=orange!80!black] (0.225, -0.8) rectangle ++(\bufferlength/2,-\bufferwidth);
\draw[->, color=gray, very thin] (0.3, -0.75) -- +(0, 0.6);
\node at (2,0.8) {\scriptsize NACK};
\node at (3,0.8) {\scriptsize NACK};
\node at (4,0.8) {\scriptsize ACK};

\node at (-1.2,\buffery-0.1) (a) {\scriptsize Buffer status};
\node[below = 0.55cm of a.west, anchor = west]  {\scriptsize Packet arrival};



\draw[->] (0,\preempy) -- +(5.5,0) node[anchor=west] (timeb) {\scriptsize Channel};
\node[below = 0.25cm of timeb.west, anchor = west]  {\scriptsize uses};

\foreach \x in {0,...,25} {
  \draw[-,>=stealth] (\x*0.2,\preempy-0.05) -- +(0,0.1);
}

\foreach \x in {0,5,10,15,20, 25} {
  \draw[-, thick] (\x*0.2,\preempy-0.1) -- +(0,0.2);
}

\draw[draw=black, fill=orange!80!black] (5*0.2,\preempy+0.3) rectangle ++(1,0.3);
\draw[draw=black, fill=orange!80!black] (10*0.2,\preempy+0.3) rectangle ++(1,0.3);

\foreach \x in {15,20} {
  \draw[draw=black, fill=green!80!black] (\x*0.2,\preempy+0.3) rectangle ++(1,0.3);
}

\draw (-0.3, \preempy+ \buffery) -- ++(\bufferlength,0) -- ++(0,-\bufferwidth) -- ++(-\bufferlength,0);
\draw (-0.15, \preempy+\buffery) -- +(0,-\bufferwidth);
\draw (0.7,\preempy+ \buffery) -- ++(\bufferlength,0) -- ++(0,-\bufferwidth) -- ++(-\bufferlength,0);
\draw[draw=black, fill=orange!80!black] (0.85,\preempy+ \buffery) rectangle ++(\bufferlength/2,-\bufferwidth);
\draw (1.7, \preempy+ \buffery) -- ++(\bufferlength,0) -- ++(0,-\bufferwidth) -- ++(-\bufferlength,0);
\draw[draw=black, fill=orange!80!black] (1.85,\preempy+ \buffery) rectangle ++(\bufferlength/2,-\bufferwidth);
\draw (2.7, \preempy+ \buffery) -- ++(\bufferlength,0) -- ++(0,-\bufferwidth) -- ++(-\bufferlength,0);
\draw[draw=black, fill=green!80!black] (2.85, \preempy+\buffery) rectangle ++(\bufferlength/2,-\bufferwidth);
\draw (3.7, \preempy+\buffery) -- ++(\bufferlength,0) -- ++(0,-\bufferwidth) -- ++(-\bufferlength,0);
\draw[draw=black, fill=green!80!black] (3.85,\preempy+ \buffery) rectangle ++(\bufferlength/2,-\bufferwidth);

\draw (4.7, \preempy+\buffery) -- ++(\bufferlength,0) -- ++(0,-\bufferwidth) -- ++(-\bufferlength,0);
\draw (4.85,\preempy+ \buffery) -- +(0,-\bufferwidth);

\draw[draw=black, fill=orange!80!black] (0.225, \preempy-0.8) rectangle ++(\bufferlength/2,-\bufferwidth);
\draw[->, color=gray, very thin] (0.3, \preempy-0.75) -- +(0, 0.6);

\draw[draw=black, fill=green!80!black] (2.225,\preempy -0.8) rectangle ++(\bufferlength/2,-\bufferwidth);
\draw[->, color=gray, very thin] (2.3,\preempy -0.75) -- +(0, 0.6);

\node at (2,\preempy+0.8) {\scriptsize NACK};
\node at (3,\preempy+0.8) {\scriptsize NACK};
\node at (4,\preempy+0.8) {\scriptsize NACK};
\node at (5,\preempy+0.8) {\scriptsize ACK};

\node at (-1.2,\preempy+\buffery-0.1) (a) {\scriptsize Buffer status};
\node[below = 0.55cm of a.west, anchor = west]  {\scriptsize Packet arrival};

\end{tikzpicture}
          \caption{Top figure: a packet is successfully delivered after three ARQ rounds.
          Bottom figure: after two unsuccessful ARQ rounds, the packet under service is discarded because a new, fresher packet has become available at the transmitter (preemption in service). The new packet is successfully delivered after two ARQ rounds.
         }
        \label{fig:policy}
\end{figure}
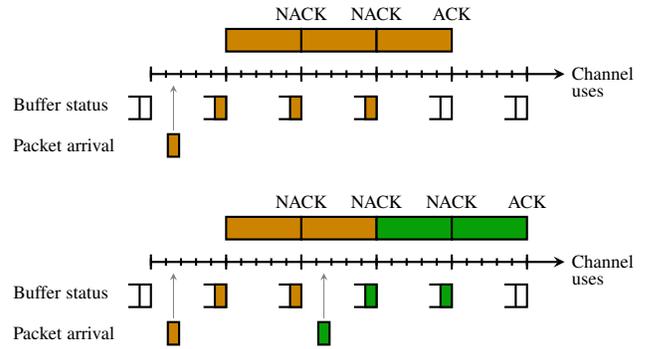

As in~\cite{devassy18-12a}, we consider an LCFS queue discipline with system capacity $1$ and preemption in service.
An arriving packet is stored in the queue provided that the queue is empty.
Specifically, upon the arrival of a new information packet, if the system is busy, the packet under service is discarded, and the transmission of the new packet commences---see Fig.~\ref{fig:policy}.

The service process operates as follows: each information packet is mapped into a coded packet of length $n$.
Hence, the transmission rate is $R=k/n$.
We assume that time is organized into frames of length $n$ channel uses, and that a new coded packet is transmitted starting from the first available frame using a simple \gls{arq} protocol.
Specifically, if the receiver decodes the coded packet successfully, it sends an ACK message to the transmitter, which removes the packet from the queue.
If the packet is not decoded successfully, the receiver sends a NACK and the coded packet is retransmitted.
For simplicity, we adopt the common assumption of perfect error detection at the receiver and instantaneous error-free ACK/NACK transmission.

As in~\cite{ferrante18}, the propagation channel is modeled as a spatially white \gls{mimo} Rayleigh memoryless block-fading channel with $m\sub{t}$ transmit and $m\sub{r}$ receive antennas.
Specifically, we assume that each packet spans $\ell$ coherence blocks of size $n\sub{c}$ channel uses, so that $n=\ell n\sub{c}$.
This model is relevant for multicarrier-based systems like 5G in which one can exploit different degrees of frequency diversity (captured in our model by the parameter $\ell$) by spacing the resource blocks sufficiently far apart in frequency.
Obviously, for a fixed packet size $n$, the larger $\ell$, the smaller the size $n\sub{c}$ of the coherence block, and, hence, the larger the channel estimation overhead in the no-\gls{csi} case of interest in this study.
The signal received during coherence block $j\in \lrbo{1,\dots,\ell}$ is given by
\begin{IEEEeqnarray}{rCL}\label{eq:io-relation}
  \rmatY_{j}=\rmatH_{j}\rmatX_{j}+\rmatW_{j}
\end{IEEEeqnarray}
where $\rmatX_{j}\in\complexset^{m\sub{t}\times n\sub{c}}$ is the channel input; $\rmatY_{j}\in\complexset^{m\sub{r}\times n\sub{c}}$ is the channel output; $\rmatH_{j}\in\complexset^{m\sub{r}\times m\sub{t}}$ is the fading matrix, whose entries are drawn independently from a $\jpg(0,1)$ distribution; finally, $\rmatW_{j} \in \complexset^{m\sub{r}\times n\sub{c}}$, whose entries are again drawn independently from a $\jpg(0,1)$ distribution, models the additive Gaussian noise.
We assume that the matrices $\rmatH_{j}$ and $\rmatW_{j}$ are mutually independent and take independent values across coherence blocks.
Furthermore, we assume that the probability distribution of $(\rmatH_{j},\rmatW_{j})$  does not depend on the input signal.
Finally, no a priori knowledge of the realization of the fading matrices $\{\rmatH_{j}\}$ is available at the transmitter and the receiver.
This implies that no-\gls{csi} is available.

To enable channel estimation at the receiver, we consider pilot-assisted transmission, according to which pilot symbols known by the receiver are embedded in the signal to be transmitted in each block.
Specifically, we assume that $\rmatX_{j}=[\rmatX_{j}^{(\mathrm{p})} \rmatX_{j}^{(\mathrm{d})}]$ where $\rmatX_{j}^{(\mathrm{p})} \in \complexset^{m\sub{t}\times n\sub{p}}$, with $1\leq n\sub{p} < n\sub{c}$, is a deterministic matrix containing orthogonal sequences of $n\sub{p}$ pilot symbols in each row, and $\rmatX_{j}^{(\mathrm{d})}\in \complexset^{m\sub{t}\times (n\sub{c}-n\sub{p})}$ contains the data symbols.

{Although our framework is general and holds for arbitrary input distributions, we shall assume for simplicity that both pilot and data symbols are drawn from a QPSK constellation and that each symbol has power $\rho/m\sub{t}$.}
Note that since the noise in~\eqref{eq:io-relation} has unit variance, we can interpret $\rho$ as the SNR.

We assume that the receiver uses the pilot symbols to compute a maximum likelihood estimate $\widehat{\rmatH}_j$ of the fading matrix $\rmatH_j$, $j=1,\dots,\ell$.
The estimate is then used to perform mismatched scaled nearest neighbor decoding to determine the transmitted codeword.
Specifically, let $\rmatY_{j}=[\rmatY_{j}^{(\mathrm{p})} \rmatY_{j}^{(\mathrm{d})}]$ where $\rmatY_{j}^{(\mathrm{p})}$ and $\rmatY_{j}^{(\mathrm{d})}$ contain the received symbols corresponding to the transmitted pilot and data symbols, respectively.
{As in~\cite{Ostman18-10}, the receiver} computes $\widehat{\rmatH}_j$ as
\begin{IEEEeqnarray}{rCL}\label{eq:ml-channel-estimate}
    \widehat{\rmatH}_j=\frac{m\sub{t}}{\rho n\sub{p}} \rmatY_{j}^{(\mathrm{p})} \left(\rmatX_{j}^{(\mathrm{p})}\right)^H.
\end{IEEEeqnarray}
The decoder then chooses the codeword $\{\rmatX_j\}_{j=1}^{\ell}$ that minimizes the nearest-neighbor decoding metric
$\prod_{j=1}^{\ell}q(\rmatX_{j},\rmatY_{j})$ where
\begin{IEEEeqnarray}{rCL}\label{eq:snn-decoding-metric}
  q(\rmatX_j,\rmatY_j)=\prod_{i=1}^{n\sub{c}-n\sub{p}} \exp\lefto(
  -\vecnorm{\vecy_{j,i}^{(\mathrm{d})}-\widehat{\rmatH}_j\vecx_{j,i}^{(\mathrm{d})}}^2
  \right).
\end{IEEEeqnarray}
Here, $\vecy_{j,i}^{(\mathrm{d})}$ and $\vecx_{j,i}^{(\mathrm{d})}$ denote the $i$th column of the matrices $\rmatY_{j}^{(\mathrm{d})}$ and $\rmatX_{j}^{(\mathrm{d})}$, respectively.

\section{Packet Error Probability Characterization through Finite-Blocklength Information Theory} \label{sec:packet_error_probability_characterization_through_finite_blocklength_information_theory}
As we shall discuss in Section~\ref{sec:peak_age_of_information}, a key ingredient to compute the PAoI violation probability is the availability of a finite-blocklength bound on the packet error probability achievable in a single packet transmission (within the \gls{arq} protocol), using the modulation scheme and the decoder described in Section~\ref{sec:system_model}, for a given number of information bits, transmit and receive antennas, diversity branches, size of the coherence block, and SNR.

Next, we provide one such finite-blocklength bound, which is based on the random-coding union bound with parameter~$s$~\cite{martinez11-02a}---an adaptation of the random-coding union bound~\cite[Thm.~16]{polyanskiy10-05a} to the case of mismatch decoding.
This bound is stated in the following theorem, which follows from~\cite[Thm.~1]{ferrante18} (see also~\cite[Thm.~3]{Ostman18-10}).
\begin{thm}\label{thm:rcus}
  Fix an integer $1\leq n\sub{p}<n\sub{c}$ and a real number $\alpha\geq 0$.
  Let the generalized information density be defined as
  \begin{IEEEeqnarray}{rCL}\label{eq:generalized-info-density}
    \imath_\alpha(\rmatX_j,\rmatY_j)=\log\frac{q(\rmatX_j,\rmatY_j)^\alpha}{\Ex{
    \overline{\rmatX}_j}{q(\overline{\rmatX}_j,\rmatY_j)^\alpha}}.
  \end{IEEEeqnarray}
  Here, $\rmatX_j$ contains all symbols transmitted in the $j$th block, in particular any symbol that may be used for pilot or data is contained in $\rmatX_j$. 
  Furthermore, the entries of $\rmatX_j$ are drawn independently and uniformly from the QPSK alphabet.
  The matrix $\rmatY_j$ contains the corresponding received symbols.
  $\overline{\rmatX}_j$ is distributed as $\rmatX_j$, and is independent of both $\rmatX_j$ and $\rmatY_j$.
The average packet error probability $\epsilon$ achievable using the modulation and decoding strategy described in Section~\ref{sec:system_model} is upper-bounded as
\begin{multline}
  \epsilon\leq \bar{\epsilon}\\=\Ex{}{\exp\biggl(-\biggl[\biggl(\sum_{j=1}^\ell   \imath_{\alpha}(\rmatX_j,\rmatY_j)\biggr)-\log(2^k-1)\biggr]^{+} \biggr)}.\label{eq:rcus}
\end{multline}

\end{thm}

\section{Peak Age of Information and peak-age violation probability}
\label{sec:peak_age_of_information}
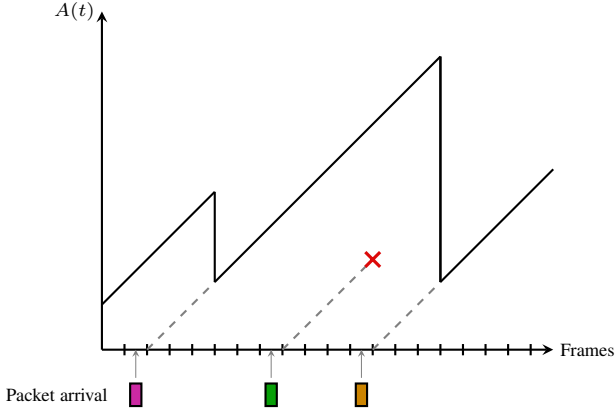
\begin{figure}[t]
        \centering
          \def \bufferlength{0.2}
\def \bufferwidth{0.3}
\def \buffery{-0.3}

\begin{tikzpicture}[thick,scale=1.5, every node/.style={scale=1}]

\tikzset{
block/.style = {draw, fill=white, rectangle, minimum height=2em, minimum width=3em},
bufferblock/.style = {draw, fill=white, rectangle, minimum height=2cm, minimum width=1cm},
tmp/.style  = {coordinate},
sum/.style= {draw, fill=white, circle, node distance=1cm},
input/.style = {coordinate},
output/.style= {coordinate},
pinstyle/.style = {pin edge={to-,thin,black}},
queuei/.pic={
  \draw (0,0) -- ++(2cm,0) -- ++(0,-0.5cm) -- ++(-2cm,0);
   \foreach \c [count =\x from 1] in {$\randvecy_1$,$\cdots$,$\randvecy_{v}$}
   {
    \draw ([xshift=-\x*15pt]2cm,0) -- ++(0,-0.5cm) {};
    \node[text centered] at ([xshift=-\x*15pt]2.3cm,-0.25cm) {\c};
   }
   \node[above] at (1cm,0) {Buffer};
  },
  queuel/.pic={
  \draw (2cm,0) -- ++(-2cm,0) -- ++(0,-0.5cm) -- ++(2cm,0);
   \foreach \c [count =\x from 1] in {,,}
   {
    \draw ([xshift=-\x*15pt]2.1cm,0) -- ++(0,-0.5cm) {};
   }
  },
  bsc/.pic={
     \node[circle,fill=blue,inner sep=1pt,minimum size=1pt] (A1) at (0,0.5) {} ;
      \node[circle,fill=blue,inner sep=1pt,minimum size=1pt] (A2)  at (0,0) {} ;
     \node[circle,fill=blue,inner sep=1pt,minimum size=1pt] (B1) at (0.5,0.5) {} ;
      \node[circle,fill=blue,inner sep=1pt,minimum size=1pt] (B2)  at (0.5,0) {};
      \draw[-]  (A1)--(B1) node [midway, above, sloped]  {};
      \draw[-]  (A1)--(B2) node at (1,2.4) {};
      \draw[-]  (A2)--(B1) node  at (1,0.6) {};
      \draw[-]  (A2)--(B2) node [midway, below, sloped]  {};
  },
  cross/.style={very thick,cross out, draw=red, minimum size=2*(#1-\pgflinewidth), inner sep=1pt, outer sep=1pt},
  cross/.default={3pt}
}

\draw[->] (0,0) -- +(4,0) node[anchor=west] {\scriptsize Frames };

\draw[->] (0,0) -- +(0,3) node[anchor=east] {\scriptsize $A(t)$};

\foreach \x in {1,...,19} {
  \draw[-,>=stealth] (\x*0.2,-0.05) -- +(0,0.1);
}

\node at (-0.4,-0.4)  {\scriptsize Packet arrival};

\draw[-] (0,0.4) -- (1,1.4) node[anchor=east] {};

\draw[draw=black, fill=pink!80!black] (0.25, -0.5) rectangle ++(\bufferlength/2,0.2);
\draw[->, color=gray, very thin] (0.3, -0.25) -- +(0, 0.25);

\draw[-, dashed, color=gray] (0.4,0) -- (1,0.6) node[anchor=east] {};
\draw[-] (1,1.4) -- (1,0.6) node[anchor=east] {};

\draw[-] (1,0.6) -- (3,2.6) node[anchor=east] {};

\draw[draw=black, fill=green!80!black] (1.45, -0.5) rectangle ++(\bufferlength/2,0.2);
\draw[->, color=gray, very thin] (1.4 + 0.1 , -0.25) -- +(0, 0.25);

\draw[-, dashed, , color=gray] (1.6,0) -- (2.4,0.8) node[anchor=east] {};
\draw (2.4,0.8) node[cross] {};

\draw[draw=black, fill=orange!80!black] (2.25, -0.5) rectangle ++(\bufferlength/2,0.2);
\draw[->, color=gray, very thin] (2.2 + 0.1 , -0.25) -- +(0, 0.25);
\draw[-, dashed, color=gray] (2.4,0) -- (3,0.6) node[anchor=east] {};
\draw[-] (3,2.6) -- (3,0.6) node[anchor=east] {};
\draw[-] (3,2.6) -- (3,0.6) node[anchor=east] {};
\draw[-] (3,0.6) -- (4,1.6) node[anchor=east] {};

\end{tikzpicture}
          \caption{Evolution of the age of information for the LCFS-S policy. Note that the second packet is discarded because, during its service time, a fresher packet becomes available at the transmitter.
         }
        \label{fig:age}
\end{figure}

We assume that each information packet carries a source-encoded sample of a random process  together with the time at which the sample was taken.
As time stamp, we use the index of the frame in which the information packet entered the queue.
Let $T_m$ be the time stamp carried by the $m$th \emph{successfully delivered packet} (i.e., the packet was not discarded because of preemption in service).
Let $D_m$ be the service time of the $m$th successfully delivered packet.
Then, the index of the most recent packet received at the destination by time frame $t$ is
\begin{IEEEeqnarray}{rCL}\label{eq:index-most-recent-packet}
  \widehat{m}(t)=\max\{m\sothat T_m + D_m \leq t\}
\end{IEEEeqnarray}
and the corresponding time stamp is $T_{\widehat{m}(t)}$.
The age of information is
the discrete-time random process $A(t)=t-T_{\widehat{m}(t)}$.
The PAoI $A_m$ is the value of $A(t)$ just before the $m$th successfully delivered packet is received.
An example of the evolution of $A(t)$ for the LCFS-S policy described in Section~\ref{sec:system_model} is depicted in Fig.~\ref{fig:age}.

Let $A$ be the steady-state PAoI.
Note that, for our specific setup, the steady-state PAoI distribution exists since the packet interarrival times and the service times follow a geometric distribution. 
For a target number of channel uses $a$, we define the PAoI violation probability $P\sub{av}(a)$ as the probability that the PAoI at steady state $A$ exceeds $a/n${~\cite[Sec. V]{devassy18-12a}}, i.e.,
 \begin{IEEEeqnarray}{rCL}\label{eq:PAoI-violation-prob}
   P\sub{av}(a)=\Prob\lefto[A\geq \frac{a}{n}\right].
 \end{IEEEeqnarray}
 Note that measuring the threshold $a$ in channel uses rather than in frames allows one to assess the impact on~\eqref{eq:PAoI-violation-prob} of different choices of the frame size.
 As in~\cite{Costa16-04,inoue17-06a,devassy18-12a}, it is convenient to characterize $P\sub{av}(a)$ indirectly through the \gls{pgf} $G_{A}(s)$ of $A$. The PAoI violation probability can then be determined through an inverse transform, whose numerical evaluation can be performed as discussed in~\cite[App.~B]{devassy18-12a}.

For the case of the LCFS-S policy described in Section~\ref{sec:system_model}, and for simple \gls{arq} transmission with single-transmission packet error probability $\bar{\epsilon}$ given in~\eqref{eq:rcus}, the \gls{pgf} of $A$ is given in the following theorem, which is a simple adaptation of~\cite[Thm.~8]{devassy18-12a}.
\begin{thm}\label{thm:pgf-peak-age}
  %
  When packets arriving at average rate $\lambda$ are served with the LCFS-S policy using simple \gls{arq}, the steady-state \gls{pgf} $G_A(s)$ of the PAoI is given by
  \begin{IEEEeqnarray}{rCL}\label{eq:pgf-age}
    G_{A}(s)=G_{T^{(0)}}(s)\frac{p\, G_{H^{(0)}}(s)}{1-(1-p)G_{H^{(0)}}(s)}
  \end{IEEEeqnarray}
  where
  \begin{equation}\label{eq:pgf-cond-serv-time}
    G_{H^{(0)}}(s)=\frac{\bigl[(1-\bar{\epsilon}(1-\lambda)^n)\bigr]s}{1-\bar{\epsilon}(1-\lambda)^n s}
  \end{equation}
  is the \gls{pgf} of the conditional service time $H^{(0)}$ given that the service is not preempted,
  \begin{equation}\label{eq:prob_p}
    p=\frac{1-\bar{\epsilon}}{1-\bar{\epsilon}(1-\lambda)^n}
  \end{equation}
  is the probability that no preemption in service occurs,
  \begin{equation}\label{eq:pgf-time-elapsed-packet-ok}
    G_{T^{(0)}}(s)=\frac{G_T(s)-(1-p)G_{H^{(0)}}(s)}{p}
  \end{equation}
  is the \gls{pgf} of the amount of time $T^{(0)}$ elapsed between two packets, given that the first one is delivered successfully (no preemption), and
  \begin{equation}\label{eq:eq-pgf-time-elapsed}
    G_{T}(s)=\frac{\bigl[1-(1-\lambda)^n\bigr]s}{1-(1-\lambda)^n s}
  \end{equation}
  is the \gls{pgf} of the amount of time $T$ elapsed between the arrival of two packets.
\end{thm}

In the limit $\lambda\to 1$, the PAoI violation probability $\Pv$ admits the following simple expression~\cite[Sec. VI]{devassy18-12a}
\begin{IEEEeqnarray}{rCl}
  \lim_{\lambda\rightarrow 1}\Pv &=& \prob{H \geq a/n - 1} \label{eq:lim_Pv}
\end{IEEEeqnarray}
where $H \sim \mathrm{Geom}\lro{1-\bar{\epsilon}}$.

\section{Numerical Results}

In this section, we characterize the dependence of the PAoI violation probability $\Pv$ on the average packet arrival rate $\lambda$ and on the parameters of the underlying wireless fading channel, such as the SNR, the number of available diversity branches, and of transmit and receive antennas.
Such parameters influence $\Pv$ through the single-transmission packet error probability~$\bar{\epsilon}$ in~\eqref{eq:rcus}.

Throughout, we assume that the number of information bits per packet is $k=30$ and that a coded packet has size $n=100$.
This yields a single-transmission rate $R = 0.3$ bits per channel use.

\begin{figure}[t]
    \centering
        \begin{tikzpicture}
  \pgfplotsset{
      scaled y ticks = false,
      width=\fwidth*0.45,
      height=\fwidth*0.4,
       title style={yshift=-6pt,}
  }
 \begin{axis}
     [
      xlabel={$\bar{\epsilon}$},
      ylabel = {$\lim_{\lambda \rightarrow 1} \Pv$},
      grid=both,
      ymin = 1e-10,
      ymax =1,
      xmin = 1e-3,
      xmax = 1,
      ymode = log,
      xmode = log,
      xticklabels={{$10^{-4}$}, {$10^{-3}$}, {$10^{-2}$}, {$10^{-1}$},{$10^{0}$}},
       xtick={0.0001, 0.001, 0.01, 0.1, 1},
       legend style={at={(0.1,0.8)},anchor=west}
     ]

 \addplot [color=blue!80!black, line width=\linew] table [y index={1}, x index = {0}, col sep=comma] {./Simulation_data/limiting_Pav_vs_epsilon.csv};\addlegendentry{$a=400$}
 \addplot [color=red!80!black, line width=\linew, dashed] table [y index={3}, x index = {0}, col sep=comma] {./Simulation_data/limiting_Pav_vs_epsilon.csv};\addlegendentry{$a=800$}

\end{axis}

\end{tikzpicture}
        \caption{Limiting PAoI violation probability $\Pv$ as $\lambda \rightarrow 1$ for the LCFS-S policy combined with an ARQ protocol, as a function the single-transmission packet error probability $\bar{\epsilon}$.}
        \label{fig:Pav_vs_eps}
\end{figure}
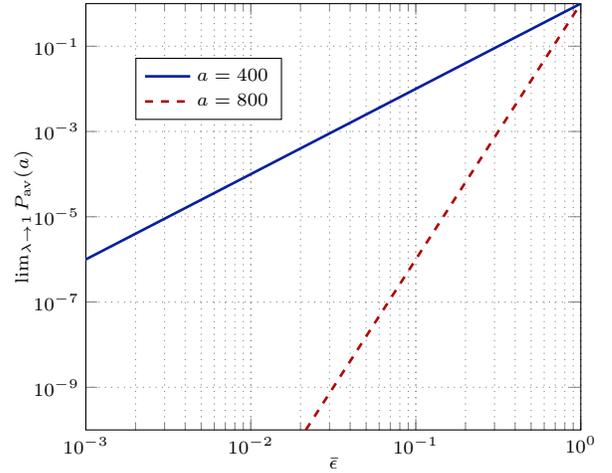

To shed light on the dependence of $\Pv$ on the packet-error probability $\bar{\epsilon}$, we first consider the asymptotic case $\lambda\rightarrow 1$, for which $\Pv$ admits the simple expression provided  in~\eqref{eq:lim_Pv}.
The value of $\lim_{\lambda\to 1}\Pv$ is illustrated in Fig.~\ref{fig:Pav_vs_eps} as a function of $\bar{\epsilon}$, for the two cases $a=400$ and $a=800$ channel uses.
As expected, a lower packet error probability $\bar{\epsilon}$ results in a lower PAoI violation probability.
The figure reveals that to achieve a PAoI violation probability not exceeding $10^{-5}$, one needs to design the physical layer so as to achieve a  packet-error probability $\bar{\epsilon}$ below $3.2\times 10^{-3}$ for the case $a=400$ channel uses, and below $1.46\times 10^{-1}$ for the case $a=800$ channel uses.

\begin{figure}[t]
    \centering
        \begin{tikzpicture}
  \pgfplotsset{
      scaled y ticks = false,
      width=\fwidth*0.45,
      height=\fwidth*0.4,
       title style={yshift=-6pt,}
  }
 \begin{axis}
     [
      xlabel={$\lambda$},
      ylabel = {$ \Pv$},
      grid=both,
      ymin = 1e-6,
      ymax =1,
      xmin = 1e-3,
      xmax = 1,
      ymode = log,
      xmode = log,
      xticklabels={{$10^{-3}$}, {$10^{-2}$}, {$10^{-1}$},{$10^{0}$}},
       xtick={ 0.001, 0.01, 0.1, 1},
       legend style={at={(axis cs: 3e-2,8e-2)},anchor=west},
       legend style={cells={align=left}}
     ]

 \addplot [color=blue!80!black, line width=\linew] table [y index={1}, x index = {0}, col sep=comma] {./Simulation_data/LCFSS_lambda_sweep_k_30_n_100_a0_400.csv};\addlegendentry{$a=400$\\ $\bar{\epsilon} = 3.2\times 10^{-3}$}
 \addplot [color=red!80!black, line width=\linew, dashed] table [y index={1}, x index = {0}, col sep=comma] {./Simulation_data/LCFSS_lambda_sweep_k_30_n_100_a0_800.csv};\addlegendentry{$a=800$\\ $\bar{\epsilon} = 1.46\times 10^{-1}$}

\end{axis}

\end{tikzpicture}
        \caption{Peak-age violation probability $\Pv$ versus $\lambda$ for the LCFS-S policy  combined with an ARQ protocol, for the two cases $a=400$, $\bar{\epsilon}=3.2\times 10^{-3}$ and $a=800$, $\bar{\epsilon}=1.46\times 10^{-1}$.}
        \label{fig:Pav_vs_lambda}
\end{figure}
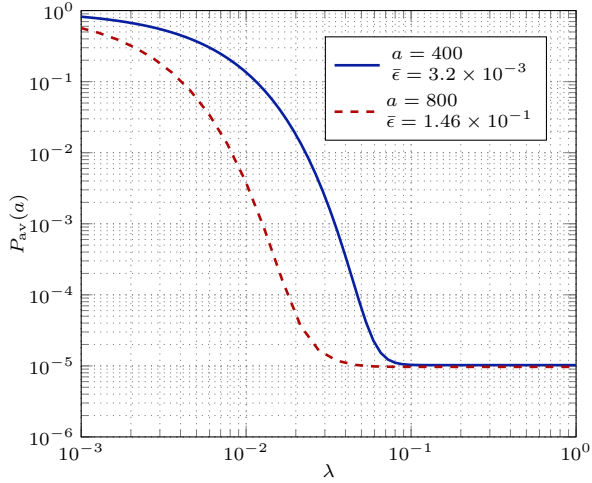

In Fig.~\ref{fig:Pav_vs_lambda}, we plot $\Pv$ as a function of $\lambda$ for the case $a=400$, $\bar{\epsilon}=3.2\times 10^{-3}$ and the case $a=800$, $\bar{\epsilon}=1.46\times 10^{-1}$.
The curves are obtained by using Theorem~\ref{thm:pgf-peak-age} together with the numerical method described in~\cite[App.~B]{devassy18-12a} to evaluate $\Pv$ from $G_A(s)$ in~\eqref{eq:pgf-age}.
As expected, both curves converge to $10^{-5}$ as $\lambda\to 1$.
Note also that an average packet arrival rate $\lambda=5\times 10^{-2}$ for the case $a=800$, and $\lambda=9\times 10^{-2}$ for the case $a=400$ is sufficient to operate close to the asymptotic limit of the PAoI violation probability.
This implies that, if the average packet arrival rate can be controlled by the system designer, it should be chosen so as not to exceed these two rates, since this would only result in an increase in packet preemption, without any gain in $\Pv$.

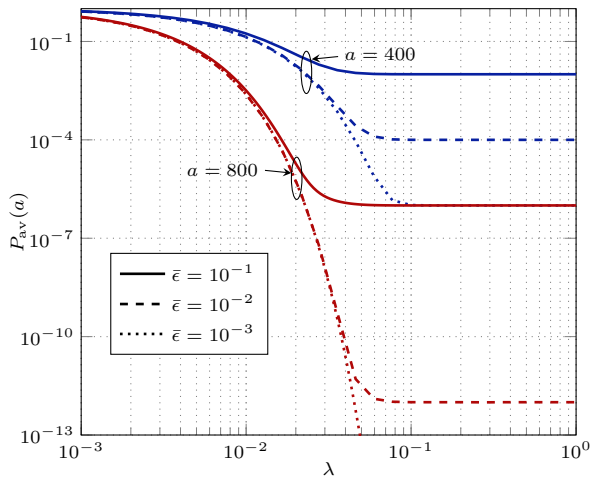
\begin{figure}[t]
    \centering
        \begin{tikzpicture}
  \pgfplotsset{
      scaled y ticks = false,
      width=\fwidth*0.45,
      height=\fwidth*0.4,
       title style={yshift=-6pt,}
  }
 \begin{axis}
     [
      xlabel={$\lambda$},
      ylabel = {$ \Pv$},
      grid=both,
      ymin = 1e-13,
      ymax =1,
      xmin = 1e-3,
      xmax = 1,
      ymode = log,
      xmode = log,
      xticklabels={{$10^{-4}$}, {$10^{-3}$}, {$10^{-2}$}, {$10^{-1}$},{$10^{0}$}},
       xtick={0.0001, 0.001, 0.01, 0.1, 1},
       legend style={at={(axis cs: 1.5e-3,1e-9)},anchor=west},
       legend image post style={black},
       legend style = {text=black}
     ]

 \addplot [color=blue!80!black, line width=\linew] table [y index={1}, x index = {0}, col sep=comma] {./Simulation_data/Vary a and eps/LCFSS_lambda_sweep_k_30_n_100_SNR_-3_a0_400_epsilon_0.1.csv}coordinate[pos=0.65](ut1);\addlegendentry{$\bar{\epsilon}=10^{-1}$}\label{pa1}
 \addplot [color=blue!80!black, line width=\linew, dashed] table [y index={1}, x index = {0}, col sep=comma] {./Simulation_data/Vary a and eps/LCFSS_lambda_sweep_k_30_n_100_SNR_-3_a0_400_epsilon_0.01.csv};\addlegendentry{$\bar{\epsilon}=10^{-2}$}\label{pa2}
 \addplot [color=blue!80!black, line width=\linew, dotted] table [y index={1}, x index = {0}, col sep=comma] {./Simulation_data/Vary a and eps/LCFSS_lambda_sweep_k_30_n_100_SNR_-3_a0_400_epsilon_0.001.csv};\addlegendentry{$\bar{\epsilon}=10^{-3}$}\label{pa3}

 \addplot [color=red!80!black, line width=\linew] table [y index={1}, x index = {0}, col sep=comma] {./Simulation_data/Vary a and eps/LCFSS_lambda_sweep_k_30_n_100_SNR_-3_a0_800_epsilon_0.1.csv}coordinate[pos=0.65](ut2);
 \addplot [color=red!80!black, line width=\linew, dashed] table [y index={1}, x index = {0}, col sep=comma] {./Simulation_data/Vary a and eps/LCFSS_lambda_sweep_k_30_n_100_SNR_-3_a0_800_epsilon_0.01.csv};
 \addplot [color=red!80!black, line width=\linew, dotted] table [y index={1}, x index = {0}, col sep=comma] {./Simulation_data/Vary a and eps/LCFSS_lambda_sweep_k_30_n_100_SNR_-3_a0_800_epsilon_0.001.csv};


\coordinate (pt2) at ($(ut1)+ (0pt,-5pt)$);
\draw[rotate=0](pt2) ellipse  (2pt and 8pt);
\coordinate (pt3) at ($(pt2)+ (2pt,5pt)$);
\coordinate (pt4) at ($(pt3)+ (+10pt,1pt)$);
\draw[<-] (pt3)--(pt4) node at ($(pt4) + (16pt,1pt)$) {$a=400$};

\coordinate (pt2) at ($(ut2)+ (0pt,-5pt)$);
\draw[rotate=0](pt2) ellipse  (2pt and 8pt);
\coordinate (pt3) at ($(pt2)+ (-2pt,2pt)$);
\coordinate (pt4) at ($(pt3)+ (-10pt,1pt)$);
\draw[<-] (pt3)--(pt4) node[anchor=east] at ($(pt4) + (0pt,0pt)$) {$a=800$};

\end{axis}

\end{tikzpicture}
        \caption{PAoI violation probability for the LCFS-S policy as a function of $\lambda$. The behavior of $\Pv$ is shown for different PAoI constraints $a$ and packet error probabilities $\bar{\epsilon}$.}
        \label{fig:Pav_vs_eps_and_a}
\end{figure}


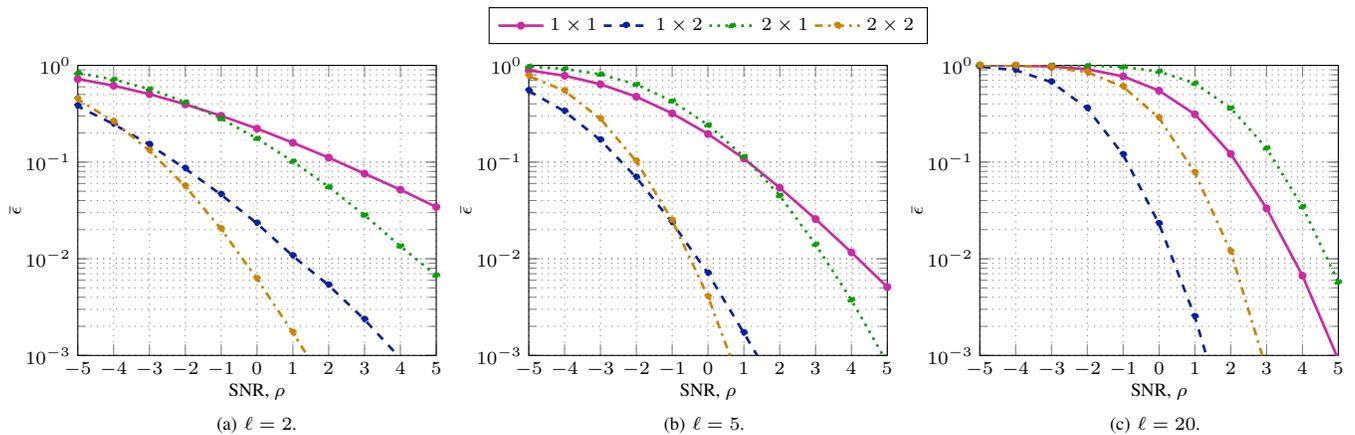
\begin{figure*}
          \centering
           \begin{tikzpicture}
\pgfplotsset{
    scaled y ticks = false,
    width=\fwidth*0.35,
    height=\fwidth*0.3,
     title style={yshift=-6pt,}
}
    \begin{groupplot}[
        group style={
        group name = my plots,
        group size=3 by 1,
        vertical sep=45pt,
        horizontal sep=35pt
        },
    ]
    \nextgroupplot[
    xlabel={SNR, $\rho$},
    ylabel = {$\bar{\epsilon}$},
    grid=both,
    ymin = 1e-3,
    ymax =1,
    xmin = -5,
    xmax = 5,
    ymode = log,
    xticklabels={{$-5$}, {$-4$}, {$-3$}, {$-2$},{$-1$}, {$0$}, {$1$}, {$2$},{$3$}, {$4$}, {$5$}},
      xtick={-5,-4,-3,-2,-1,0,1,2,3,4,5},
    ]

  \addplot [ color=pink!80!black, mark=*,mark size = 1,line width=\linew, forget plot] table [y index={1}, x index = {0}, col sep=comma] {./Simulation_data/epsilon versus SNR/QPSK_RCUs_k_30_n_100_L_2_MtMr_1x1.csv};

  \addplot [ color=blue!80!black, mark=*,mark size = 1,line width=\linew, dashed, forget plot] table [y index={1}, x index = {0}, col sep=comma] {./Simulation_data/epsilon versus SNR/QPSK_RCUs_k_30_n_100_L_2_MtMr_1x2.csv};

  \addplot [ color=green!80!black, mark=*,mark size = 1,line width=\linew, dotted, forget plot] table [y index={1}, x index = {0}, col sep=comma] {./Simulation_data/epsilon versus SNR/QPSK_RCUs_k_30_n_100_L_2_MtMr_2x1.csv};


  \addplot [ color=orange!80!black, mark=*,mark size = 1,line width=\linew, dash dot, forget plot] table [y index={1}, x index = {0}, col sep=comma] {./Simulation_data/epsilon versus SNR/QPSK_RCUs_k_30_n_100_L_2_MtMr_2x2.csv};

    \coordinate (c1) at (rel axis cs:0,1);

    \nextgroupplot[
    xlabel={SNR, $\rho$},
    ylabel = {$\bar{\epsilon}$},
    grid=both,
    ymin = 1e-3,
    ymax =1,
    xmin = -5,
    xmax = 5,
    ymode = log,
    xticklabels={{$-5$}, {$-4$}, {$-3$}, {$-2$},{$-1$}, {$0$}, {$1$}, {$2$},{$3$}, {$4$}, {$5$}},
     xtick={-5,-4,-3,-2,-1,0,1,2,3,4,5},
     legend style={at={($(0,0)+(1cm,1cm)$)},legend columns=6,fill=none,draw=black,anchor=center,align=center},
     legend to name=fredSNR
    ]
    \addplot [ color=pink!80!black, mark=*,mark size = 1,line width=\linew] table [y index={1}, x index = {0}, col sep=comma] {./Simulation_data/epsilon versus SNR/QPSK_RCUs_k_30_n_100_L_5_MtMr_1x1.csv};\addlegendentry{$1\times 1$}

    \addplot [ color=blue!80!black, mark=*,mark size = 1,line width=\linew, dashed] table [y index={1}, x index = {0}, col sep=comma] {./Simulation_data/epsilon versus SNR/QPSK_RCUs_k_30_n_100_L_5_MtMr_1x2.csv};\addlegendentry{$1\times 2$}

    \addplot [ color=green!80!black, mark=*,mark size = 1,line width=\linew, dotted] table [y index={1}, x index = {0}, col sep=comma] {./Simulation_data/epsilon versus SNR/QPSK_RCUs_k_30_n_100_L_5_MtMr_2x1.csv};\addlegendentry{$2\times 1$}

    \addplot [ color=orange!80!black, mark=*,mark size = 1,line width=\linew, dash dot] table [y index={1}, x index = {0}, col sep=comma] {./Simulation_data/epsilon versus SNR/QPSK_RCUs_k_30_n_100_L_5_MtMr_2x2.csv};\addlegendentry{$2\times 2$}


    \nextgroupplot[
    xlabel={SNR, $\rho$},
    ylabel = {$\bar{\epsilon}$},
    grid=both,
    ymin = 1e-3,
    ymax =1,
    xmin = -5,
    xmax = 5,
    ymode = log,
    xticklabels={{$-5$}, {$-4$}, {$-3$}, {$-2$},{$-1$}, {$0$}, {$1$}, {$2$},{$3$}, {$4$}, {$5$}},
     xtick={-5,-4,-3,-2,-1,0,1,2,3,4,5},
    ]
    \addplot [ color=pink!80!black, mark=*,mark size = 1,line width=\linew] table [y index={1}, x index = {0}, col sep=comma] {./Simulation_data/epsilon versus SNR/QPSK_RCUs_k_30_n_100_L_20_MtMr_1x1.csv};

    \addplot [ color=blue!80!black, mark=*,mark size = 1,line width=\linew, dashed] table [y index={1}, x index = {0}, col sep=comma] {./Simulation_data/epsilon versus SNR/QPSK_RCUs_k_30_n_100_L_20_MtMr_1x2.csv};

    \addplot [ color=green!80!black, mark=*,mark size = 1,line width=\linew, dotted] table [y index={1}, x index = {0}, col sep=comma] {./Simulation_data/epsilon versus SNR/QPSK_RCUs_k_30_n_100_L_20_MtMr_2x1.csv};


    \addplot [ color=orange!80!black, mark=*,mark size = 1,line width=\linew, dash dot] table [y index={1}, x index = {0}, col sep=comma] {./Simulation_data/epsilon versus SNR/QPSK_RCUs_k_30_n_100_L_20_MtMr_2x2.csv};

    \coordinate (c2) at (rel axis cs:1,1);

\end{groupplot}

\coordinate (c3) at ($(c1)!.5!(c2)$);
\node[above] at (c3 |- current bounding box.north)
{\pgfplotslegendfromname{fredSNR}};

\node[text width=6cm,align=center,anchor=north,font=\footnotesize] at ([yshift=-5mm]my plots c1r1.south) {\subcaption{$\ell=2$.\label{fig:epsVsSNRa}}};
\node[text width=6cm,align=center,anchor=north,font=\footnotesize] at ([yshift=-5mm]my plots c2r1.south) {\subcaption{$\ell=5$. \label{fig:epsVsSNRb}}};
\node[text width=6cm,align=center,anchor=north,font=\footnotesize] at ([yshift=-5mm]my plots c3r1.south) {\subcaption{$\ell=20$. \label{fig:epsVsSNRc}}};
\end{tikzpicture}%
          \caption{Single-transmission packet-error probability $\bar{\epsilon}$ versus SNR $\snr$ for $k=30$ and $n=100$ and for different number of transmit and receive antennas and different number of diversity branches.}
        \label{fig:epsVsSNR}
\end{figure*}

In Fig.~\ref{fig:Pav_vs_eps_and_a}, we investigate the dependence of $\Pv$ on $\lambda$, $\bar{\epsilon}$, and $a$.
As expected, $\Pv$ decreases with $a$ for a fixed $\bar{\epsilon}$\, for any value of $\lambda$.
The relation between $\Pv$ and $\bar{\epsilon}$ for a fixed $a$ differs according to the value of the average arrival rate $\lambda$.
Specifically, when $\lambda$ is small, $\Pv$ depends weakly on $\bar{\epsilon}$.
Indeed, in this regime, the PAoI violation event is triggered by the low arrival rate of the information packets, and not by the errors at the physical layer.
On the contrary, when $\lambda$ approaches one, these errors are the main cause of PAoI violation events, and $\Pv$ becomes extremely sensitive to the value of $\bar{\epsilon}$.

From Fig.~\ref{fig:Pav_vs_eps_and_a}, one notices that to satisfy a given $\Pv$ target for a given blocklength $n$ and PAoI constraint $a$, a large enough $\lambda$ and a low enough $\bar{\epsilon}$ are required.
Next, we use Theorem~\ref{thm:rcus} to determine how to design the physical layer of the communication system described in Section~\ref{sec:system_model} to achieve $\Pv\leq 10^{-5}$ for the case $a=400$ and $a=800$, which, \emph{for sufficiently large average arrival rate} $\lambda$, requires satisfying $\bar{\epsilon}\leq 3.2\times 10^{-3}$ and   $\bar{\epsilon}\leq 1.46\times 10^{-1}$, respectively (see Fig.~\ref{fig:Pav_vs_lambda}).

In Fig.~\ref{fig:epsVsSNR}, we plot the  packet-error probability as a function of the SNR $\rho$.
We consider three scenarios: the case in which a codeword spans a small ($\ell=2$, $n\sub{c}=50$), a moderate ($\ell=5$, $n\sub{c}=20$), and a large ($\ell=20$, $n\sub{c}=5$) number of diversity branches.
For a system bandwidth of $20$ MHz, these three scenarios correspond to a channel coherence bandwidth of $10$ MHz, $4$ MHz, and $1$ MHz, respectively, which is in line with the channel models used in cellular-system standardization activities~\cite{3GPP-TR-38.901}.
We also consider four antenna configurations: $1\times 1$, $1\times 2$, $2\times 1$, and $2\times 2$.
The results reported in  Fig.~\ref{fig:epsVsSNR} are obtained by optimizing $\bar{\epsilon}$ in~\eqref{eq:rcus} over the number $n\sub{p}$ of pilot symbols and over the parameter $\alpha$ in~\eqref{eq:generalized-info-density}.

These figures allow us to determine the best antenna configuration and the corresponding SNR value required to achieve the target error probabilities of\, $\bar{\epsilon}\leq 3.2\times 10^{-3}$ and\,   $\bar{\epsilon}\leq 1.46\times 10^{-1}$ as a function of the number of diversity branches available.

\begin{table}[tb]\centering
\renewcommand{\arraystretch}{1.5}
\caption{Optimal parameters in different fading scenarios to achieve $\bar{\epsilon} = 1.46\times 10^{-1} $ (required for $P\sub{av}(800) < 10^{-5}$ for a sufficiently large $\lambda$).}
\label{tab:opt_vals_a800}
\begin{tabular}{C{0.95cm}|C{2cm}C{2cm}C{2cm}}
 			& $m\sub{t} \times m\sub{r}$		& $\rho$ & $\np$ \\
      \midrule
\rowcolor{blue!30} $\ell=2$			&  $2\times 2$   			& $-3$ dB  & $14$\\
$\ell=5$			& $1\times 2$			&  $ -2.75$ dB & $6$ \\
$\ell=20$		& $1\times 2$ 			&  $ -1$ dB & $2$\\
\midrule
\end{tabular}
\end{table}

\begin{table}[tb]\centering
\caption{Optimal parameters in different fading scenarios to achieve $\bar{\epsilon} = 3.2\times 10^{-3}$ (required for $P\sub{av}(400) < 10^{-5}$ for a sufficiently large $\lambda$).}
\label{tab:opt_vals_a400}
\renewcommand{\arraystretch}{1.5}
\begin{tabular}{C{0.95cm}|C{2cm}C{2cm}C{2cm}}
 			& $m\sub{t} \times m\sub{r}$		& $\rho$ & $\np$ \\
      \midrule
$\ell=2$			&  $2\times 2$   			& $0.25$ dB  & $15$\\
\rowcolor{blue!30} $\ell=5$			& $2\times 2$			&  $0$ dB & $6$ \\
$\ell=20$		& $1\times 2$ 			&  $0.75$ dB & $2$\\
\midrule
\end{tabular}
\end{table}

The optimal parameters of the physical layer are summarized in Table~\ref{tab:opt_vals_a800} and~\ref{tab:opt_vals_a400}, where we also report the optimum number of pilot symbols per coherence block $n\sub{p}$.
We note that for the case $\bar{\epsilon}=1.46\times10^{-1}$, which corresponds to $a=800$, the lowest required SNR $\rho$ is achieved when $\ell=2$ and with a $2\times 2$ antenna configuration.
For the case $\bar{\epsilon}=3.2\times10^{-3}$, which corresponds to $a=400$, the lowest required $\rho$ is achieved when $\ell=5$ and with a $2\times 2$ antenna configuration.
A smaller $\ell$ yields the best performance for $a=800$ because the large target peak age allows for more ARQ rounds, which in turns alleviate the requirements on the packet reliability during a single transmission.
On the contrary, the higher reliability required for the case $a=400$ makes it important to exploit diversity.
It is also evident from our results that using multiple antennas at the receiver is beneficial to minimize the SNR $\rho$ required not to exceed the target PAoI violation probability.
For example, for the case $\ell=20$ and $a=400$, a $1\times 2$ antenna configuration yields a target SNR of $\rho=0.75\dB$, whereas a $1\times 1$ antenna configuration results in a target SNR of $\rho=4\dB$.
Note also that when the number of diversity branches is large and the coherence block is small (see Fig.~\ref{fig:epsVsSNRc}), the $1\times 2$ antenna configuration outperforms the $2\times 2$ antenna configuration, since it requires fewer pilot symbols to estimate the fading channel.

\section{Conclusions}
We presented a characterization of the PAoI violation probability in a wireless communication system employing \gls{arq} and an LCFS packet management policy with system capacity $1$ and preemption in service.
In contrast to previous studies in the literature, the transmission of the information packets at the physical layer is modeled in detail by using a MIMO Rayleigh block-fading channel model.
Furthermore, the packet-error probability achievable over this channel for a given packet size and transmission rate is characterized using a bound from finite-blocklength information theory.

As shown in Figs.~\ref{fig:Pav_vs_lambda} and~\ref{fig:Pav_vs_eps_and_a}, our analysis allows one to determine the packet-error probability to target in order not to exceed a given PAoI violation probability.
Furthermore, as illustrated in Table~\ref{tab:opt_vals_a800} and~\ref{tab:opt_vals_a400}, one can derive from this target probability concrete insights on the design of the physical layer of the wireless communication system, such as the SNR at which to operate, the number of transmit and receive antennas to use, and the number of pilot symbols to transmit in order to estimate the channel.

The analysis presented in this paper can be readily extended to more general fading models.
Further generalizations include the characterization of the impact of interferers, the overhead due to packet detection, extensions to hybrid \gls{arq}, and packet generation at will.

\bibliographystyle{IEEEtran}
\bibliography{./Inputs/references}
\end{document}